\documentclass[secnumarabic,superscriptaddress,amssymb, nobibnotes, aps, twocolumn, prl]{revtex4-2}

\usepackage{graphicx}  
\graphicspath{{figure/}}
\usepackage{dcolumn}   
\usepackage{bm}        
\usepackage{amssymb}   
\usepackage{amsmath}   
\usepackage{amsthm}    
\usepackage{xcolor}    
\usepackage{tikz}      
\usepackage{ifthen}    
\usepackage{multirow}  
\usepackage{tabu}      
\usepackage{soul}      
\usepackage{subfigure}
\usepackage{xspace}
\usepackage{float}
\usepackage{afterpage}
\usepackage{xr-hyper}
\usepackage[unicode=true,
  linktocpage,
  linkbordercolor={0.5 0.5 1},
  citebordercolor={0.5 1 0.5},
  linkcolor=blue]{hyperref}

\begin{document}

\title{Quantum Transport in Graphene Using Surface Acoustic Wave Resonators}%

\author{Yawen Fang}
\affiliation{Laboratory of Atomic and Solid State Physics, Cornell University, Ithaca, NY, 14853, USA}
\author{Yang Xu}
\affiliation{School of Applied and Engineering Physics and Laboratory of Atomic and Solid State Physics, Cornell University, Ithaca, NY, USA}
\affiliation{Beijing National Laboratory for Condensed Matter Physics, Institute of Physics, Chinese Academy of Sciences, Beijing, China}
\author{Kaifei Kang}
\affiliation{School of Applied and Engineering Physics, Cornell University, Ithaca, NY, USA}
\author{Benyamin Davaji}
\affiliation{Department of Electrical and Computer Engineering, Northeastern University, Boston, MA, USA}
\author{Kenji Watanabe}
\affiliation{National Institute for Materials Science, Tsukuba, Japan}
\author{Takashi Taniguchi}
\affiliation{National Institute for Materials Science, Tsukuba, Japan}
\author{Amit Lal}
\affiliation{Department of Electrical and Computer Engineering, Cornell University, Ithaca, NY, USA}
\author{Kin Fai Mak}
\affiliation{School of Applied and Engineering Physics and Laboratory of Atomic and Solid State Physics, Cornell University, Ithaca, NY, USA}
\affiliation{Kavli Institute at Cornell for Nanoscale Science, Ithaca, NY, USA}
\author{Jie Shan}
\affiliation{School of Applied and Engineering Physics and Laboratory of Atomic and Solid State Physics, Cornell University, Ithaca, NY, USA}
\affiliation{Kavli Institute at Cornell for Nanoscale Science, Ithaca, NY, USA}
\author{B.~J.~Ramshaw}\email[To whom correspondence should be addressed, ]{bradramshaw@cornell.edu}
\affiliation{Laboratory of Atomic and Solid State Physics, Cornell University, Ithaca, NY, 14853, USA}
\affiliation{Kavli Institute at Cornell for Nanoscale Science, Ithaca, NY, USA}

\begin{abstract}

Surface acoustic waves (SAWs) provide a contactless method for measuring the wavevector-dependent conductivity. This technique has been used to discover emergent length scales in the fractional quantum Hall regime of traditional, semiconductor-based heterostructures. SAWs would appear to be an ideal match for van der Waals (vdW) heterostructures, but the right combination of substrate and experimental geometry to allow access to the quantum transport regime has not yet been found. We demonstrate that SAW resonant cavities fabricated on LiNbO$_3$ substrates can be used to access the quantum Hall regime of high-mobility, hexagonal boron nitride (hBN) encapsulated graphene heterostructures. Our work establishes SAW resonant cavities as a viable platform for performing contactless conductivity measurements in the quantum transport regime of vdW materials.

\end{abstract}

\maketitle


\textit{Introduction.}---Spectroscopic probes of frequency-dependent conductivity are ubiquitous in condensed matter, but analogous techniques for measuring the momentum-dependent conductivity are relatively scarce. The fundamental problem is that photons have millimeter-scale wavelengths at the energy scales relevant to emergent phenomena, but the length scales of these phenomena, such as charge density waves or moir\'e potentials, are typically nanometers to microns.

This problem is solved by replacing light with sound. Sound waves traveling across the surface---surface acoustic waves (SAWs)---of a piezoelectric material create an oscillating electric field at the sound wavelength: tens-of-microns to tens-of-nanometers at MHz to GHz frequencies. This oscillating electric field interacts with a conducting material placed in proximity to the SAW.  By measuring the resultant changes in sound velocity and attenuation, one measures the material's conductance at the SAW wavelength. This technique is contactless, bulk, measures at finite wavelength, and is compatible with low temperatures and high magnetic fields. In the sense that it is a bulk, contactless technique, SAWs can be thought of as the transport counterpart to the thermodynamic measurement of quantum capacitance \cite{RevModPhys.54.437}.

SAWs have been used to great effect in GaAs/AlGaAs heterostructures, where naturally-piezoelectric GaAs substrates generate SAWs that interact with the 2D electron gas \cite{willett1994surface}. SAWs were used to show that current-induced modification of the bubble and stripe phases is a local phenomenon \cite{friess2018current}, to measure the periodicity and energy-momentum dispersion of the stripe phase at filling factor $9/2$ \cite{kukushkin2011collective}, and to measure the Fermi surface in the composite Fermi liquid at filling factor $1/2$ \cite{willett1993experimental}.

\begin{figure*}[t!]
\includegraphics[width=2\columnwidth]{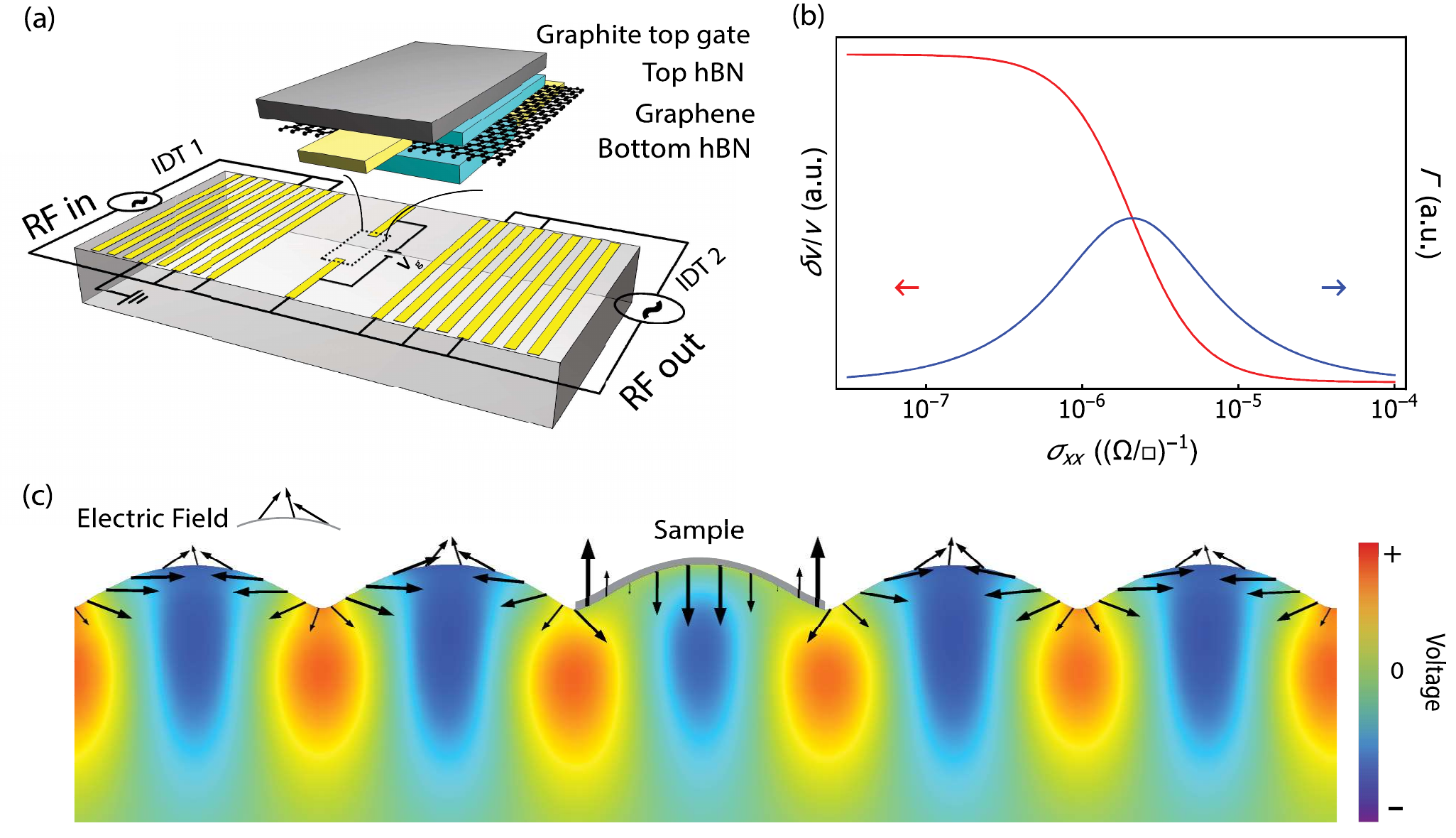}%
\caption{\textbf{Combining vdW materials with surface acoustic waves.} (a) Schematic of a SAW resonator driven by a radiofrequency (RF) source across the interdigital transducers (IDTs). A vdW heterostructure is placed in the center of the cavity and connected to gate electrodes. (b) The change in SAW velocity ($\delta v/v$) and attenuation ($\Gamma$) as a function of the longitudinal sheet conductivity of a material placed in contact with the resonator. The transition region, where the velocity change is largest and there is a peak in the attenuation, is determined by the parameters of the particular piezoelectric substrate (see SI). (c) Simulation of a SAW standing-wave resonator on lithium niobate. A conductive sample in the middle of the cavity is shown in grey. Black arrows indicate the direction and magnitude of the electric field at the surface of the substrate. In the sample region, the conducting sample screens the transverse component of the electric field: this interaction with the electric field feeds back into the stiffness of the substrate lattice, modifying the speed of sound and sound attenuation as shown in panel (b).}%
\label{fig:device}
\end{figure*}

SAWs would appear to be an ideal technique for van der Waals (vdW) materials, where it can be difficult to make electrical contact, where there are emergent and engineered length scales such as charge density wavelengths \cite{xu2020correlated} and moire periodicities \cite{cao2018correlated}, and where there are predictions for wavelength-dependent phenomena \cite{thalmeier2010surface,kumar2020fractal}. While there has been some work incorporating vdW heterostructures and SAWs \cite{lane_flip-chip_2018,yokoi2020negative}, the quantum transport regime has not been reached. 

The two main challenges are finding a suitable substrate and increasing signal size. First, a piezoelectric substrate compatible with high-mobility device fabrication and electrostatic gating must be identified. Second, even relatively large vdW heterostructures are two to four orders of magnitude smaller than the 2DEGs used in the aforementioned studies of GaAs/AlGaAs. As the SAW signal is directly proportional to the sample area, standard SAW delay-lines are not sensitive enough for quantum transport experiments on vdW heterostructures.

We have developed SAW resonant cavities on LiNbO$_3$ substrates that are compatible with high-mobility, hexagonal boron nitride (hBN)-encapsulated, graphene heterostructures. We show that the resonant cavity geometry increases signal-to-noise by two orders of magnitude over the traditional delay-line geometry. We observe strong quantum oscillations of the cavity resonance frequency in the quantum Hall regime of graphene. We comment on the present limitations of the technique and suggest future design improvements.


\textit{Background}---SAWs generate electric fields that extend approximately one wavelength above the surface of a piezoelectric substrate (\autoref{fig:device}c). This field contributes to the stiffness of the lattice and enhances the propagation velocity of the acoustic wave \cite{lewis1985rayleigh}. A conductive material placed on top of the substrate partially screens this electric field. When the conductivity material changes as a function of gate voltage, magnetic field, or temperature, the SAW velocity and attenuation coefficient also change \cite{wixforth1989surface}. These effects are plotted in \autoref{fig:device}b and are described by the relation 
\begin{equation}
\frac{\delta v_s}{v_s}-\frac{i\Gamma}{q} = \frac{K_{eff}^2}{2}\frac{1}{1+i\sigma^{\square}_{xx}(q,\omega)/\sigma_m},
\label{eq:conductivity}
\end{equation}
where $\sigma^{\square}_{xx}(q,\omega)$ is the longitudinal sheet conductance (in the direction of the SAW propagation) of the nearby material at wave vector $q$ and frequency $\omega = v_sq$, and $K_{\rm eff}^2$ is a substrate-specific coupling coefficient. $\sigma_m$ is the characteristic sheet conductance of the SAW and is given by $\sigma_m = v_s(\epsilon +\epsilon_0)$, where $\epsilon$ is the permittivity of LiNbO$_3$. $\sigma_m$ is of order $10^{-6} ~\Omega^{-1}$ for LiNbO$_3$, and $K_{\rm eff}^2 = 0.053$ \cite{vciplys1998electromechanical}. The response of the SAW to a conducting sample is maximized when the sheet conductance of the sample matches the characteristic conductance of the SAW, as shown in \autoref{fig:device}.


There are two predominant methods for generating SAWs: delay lines and resonant cavities. Most SAW experiments on GaAs/AlGaAs heterostructures use the delay-line geometry, where a pulse of sound generated at one interdigital transducer (IDT) is captured by another. Resonant cavities use the same pair of IDTs but build up a standing wave between reflectors that constrain the acoustic energy (see \autoref{fig:device}a). The signal is then effectively amplified by the quality factor ($Q$) of the cavity, which can easily be $10^3$ or higher. This suggests that the resonant cavity geometry may be preferable to the delay line geometry for small samples. 

This argument can be quantified by examining how the measured phase changes for the two geometries. For a delay line \cite{friess2017negative}, a change in sound velocity ($\delta v/v$) translates to a change in the phase ($\delta \phi/360^\circ$) of the SAW as
\begin{equation}
\frac{\delta \phi} {360^\circ} =  \frac{A_s}{\lambda \cdot W} \frac{\delta v}{v},
\label{eq:delay}
\end{equation}
where $A_s$ is the sample area, $W$ is the width of the delay line (perpendicular to the direction of propagation), and $\lambda$ is the wavelength of the SAW. For a resonant cavity with quality factor $Q$ near its resonant frequency, the phase shift is given by 
\begin{equation}
\frac{\delta \phi} {360^\circ} =  \frac{Q}{\pi} \frac{A_s}{A_c} \frac{\delta v}{v},
\label{eq:cavity}
\end{equation}
where $A_s$ and $A_c$ are the sample and cavity areas, respectively. Defining the `gain' $G$ as the cavity phase shift divided by the delay line phase shift, we obtain
\begin{equation}
G =  \frac{Q}{\pi} \frac{\lambda}{L},
\label{eq:gain}
\end{equation}
where $L$ is the length of the cavity or delay line. 

\autoref{eq:gain} suggests that, for a given $\delta v/v$ induced by a change in sample conductivity, large phase shifts can be achieved with high quality factor cavities. The factors $L$ and $\lambda$ are constrained by device fabrication: we use $\lambda = 20~\mu$m SAW cavities with an IDT separation of $L = 100 ~\mu$m. These length scales are easily achieved with standard photolithography and leave enough space to fabricate a heterostructure in the center of the cavity. These cavities regular achieve $Q$s between 1000 and 5000, producing a gain of up to 300, i.e. the phase shift is 300 times larger when the devices is operated as a cavity instead of a delay line. 


\textit{Experimental design.}---\autoref{fig:device}a shows a schematic of the two-port, Fabry-Perot SAW resonators used in this study. These resonators consist of two IDTs for excitation and detection of the SAW and two Bragg mirrors placed behind the IDTs \cite{bell1976surface}. VdW heterostructures typically require gate and grounding electrodes, as well as possible transport electrodes if desired, which we fabricated from thinner metal than the Bragg mirrors to maintain a high cavity $Q$ and to allow room for heterostructure fabrication. 
 
We fabricated our resonators on commercially-available, 128$^{\circ}$ Y-Cut, SAW-grade, black LiNbO$_3$ wafers. The device has 15 nm gold electrodes and 300 nm, electrically-isolated gold reflectors, where the mass of the gold loads the resonator surface and confines the SAWs. A second resonator is describe in the supplementary.

\begin{figure*}[t!]
\includegraphics[width=2\columnwidth]{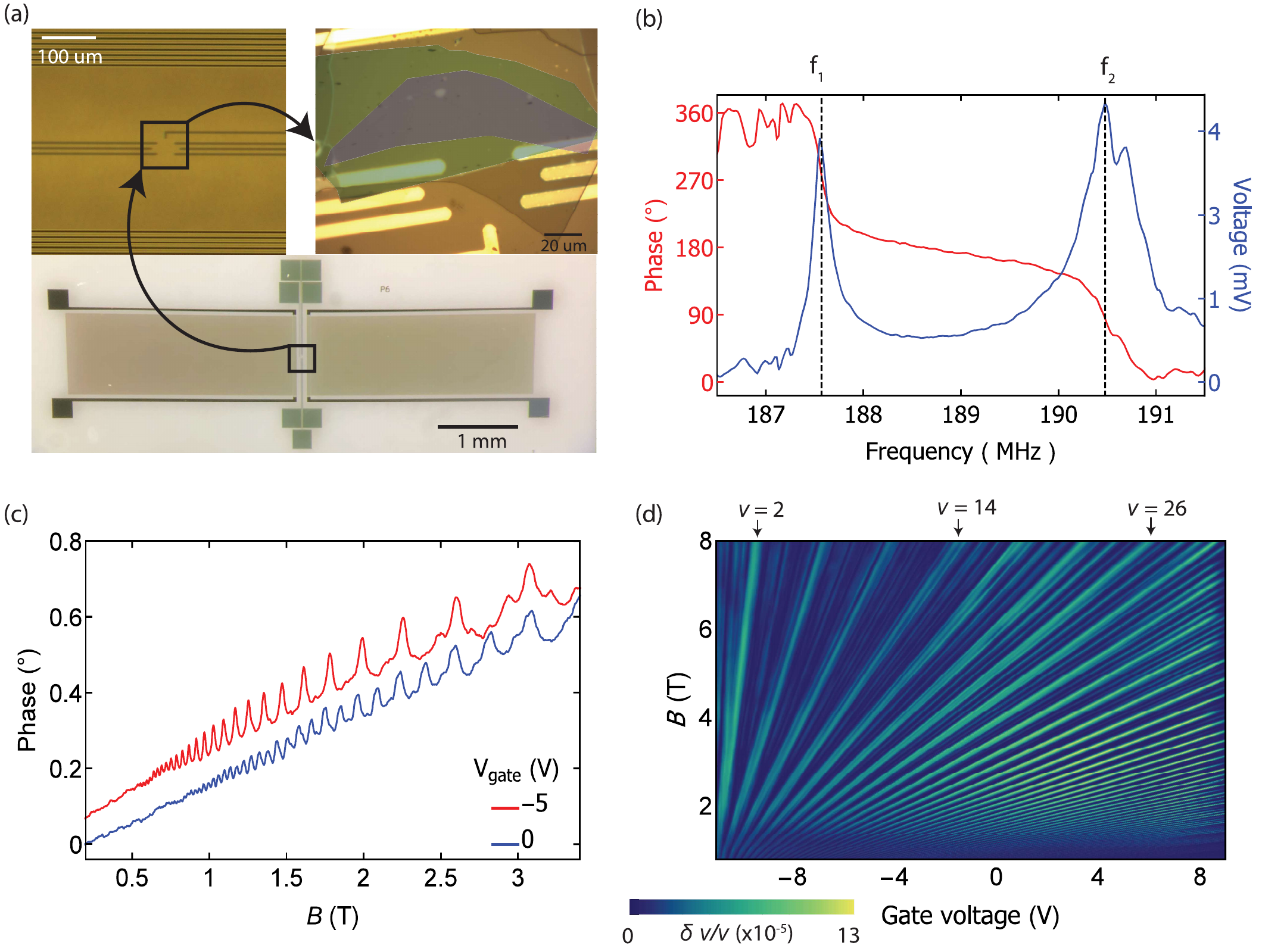}%
\caption{\textbf{Quantum transport in graphene using a surface acoustic wave resonator.} (a) Optical microscope images of the device. The lower image shows a zoomed-out view of the full SAW resonator cavity. The image on the top left shows a zoomed-in view of the center of the cavity with the detail of the pre-patterned gold electrodes without the graphene sample (the 6 transport electrodes were not used in this study). The image on the top right is a false-color image showing the graphene device in detail. The monolayer graphene is colored in light purple and the graphite top gate is colored in light green. (b) Amplitude and phase of the transmitted voltage through SAW resonant cavity as a function of frequency. Two resonances are indicated as $f_1$ and $f_2$. (c) The change in the cavity phase at fixed frequency $f_1$, as a function of magnetic field, at two different gate voltages, and at 1.7 K. Quantum oscillations are clearly visible above 0.6 tesla, and oscillations related to degeneracy-broken Landau levels are observed above 3 tesla. The data have been offset for clarity.  (d) Landau fan diagram at 1.7 K, from 1 tesla to 8 tesla. The color scale corresponds to the shift in SAW velocity, proportional to the shift in cavity phase through \autoref{eq:cavity}.}
\label{fig:S1}
\end{figure*}

The monolayer graphene devices were mechanically exfoliated and encapsulated with hBN on both sides to ensure high sample mobility.  The device shown in \autoref{fig:S1}a consists of a 37 nm bottom hBN layer, a graphene monolayer, a 64 nm top hBN layer, and a 5 nm graphite top gate. The resonator had a $Q$ of 1500 after device fabrication. 




\textit{Results.}---\autoref{fig:S1}a shows the frequency and phase response of the device. IDT1 was driven at 10 mV with a Zurich Instrument UHF lockin amplifier with both the graphene and the graphite gate grounded. The signal from IDT2 was filtered with a ZX75BP-204 band pass filter from Mini-Circuits, amplified with a AM-1571 amplifier from MITEQ, and measured with the lockin. 

Two resonances near 190 MHz, labeled $f_1$ and $f_2$, are each accompanied by a phase shift of approximately 180$^{\circ}$. Each resonance corresponds to a different integer number of wavelengths in the cavity: $f_1 - f_2 = 3$ MHz is equal to surface wave velocity, 3980 $m/s$, divided by twice the effective cavity length of 660 $\mu m$ (note that the Bragg mirrors make the effective length much longer than the physical length of 240 $\mu m$). The quality factor of the resonance at $f_1$ is 1500: we use this resonance for our measurements.


\autoref{fig:S1}c shows the change in cavity phase at the $f_1$ resonance as a function of magnetic field. Traces at two different gate voltages are shown, both taken at 1.7 K. Shubnikov-de Haas oscillations are clearly observed down to approximately 0.6 tesla, illustrating both the sensitivity of the technique and the quality of the sample. A standard Lifshitz-Kosevich analysis of the oscillation data (details in the Supplementary Information) determines the carrier densities to be $1.6 \times 10^{12}$ cm$^{-2}$ and $3.1 \times 10^{12}$ cm$^{-2}$ at -5 V and 0 V, respectively. We obtain the quantum lifetimes, $\tau_q$, using a Dingle analysis: $\tau_q\approx 0.4$ ps at -5 V and $\tau_q\approx 0.26$ ps at 0 V, corresponding to quantum mobilities greater than $10^4$ cm$^2$ V$^{-1}$s$^{-1}$. This high mobility demonstrates the compatibility of LiNbO$_3$ as a substrate material and the suitability of the cavity geometry and our fabrication procedure with high-quality vdW devices. 


Next, we measure as a function of gate voltage at fixed magnetic fields. \autoref{fig:S1}d shows a Landau fan diagram over the entire available gate voltage range and up to 8 tesla. Above 3 tesla, additional peaks emerge between the main series of peaks at filling factors $\nu = 2, 6, 10, 14,$ etc. At high gate voltage, there are clearly three intermediate peaks: these are the additional Landau levels (LLs) that emerge when Zeeman and valley degeneracy are broken, and these have been characterized extensively in high-quality transport devices \cite{young2012spin}. Note that we do not believe that any of the oscillatory features we observe are due to the graphite gate \cite{PhysRevLett.127.247702}: the 4-fold LL degeneracy we observe in \autoref{fig:S1}d is unique to graphene. The lack of oscillations from the graphite gate is because the graphite conductance is too high: the gate conductance is on the far right of \autoref{fig:device}b and any modulation of that conductance produces no significant velocity shift. A second device with slightly lower mobility, but where we can gate to the Dirac point, shows qualitatively similar features---this data is shown in the supplementary information.

\textit{Discussion.}---We have demonstrated that high-quality-factor, two-port, Fabry-Perot SAW resonators fabricated on LiNbO$_3$ can be used to measure quantum transport in graphene heterostructures. We observe degeneracy-broken LLs at less than 3 tesla and extract quantum lifetimes longer than 0.2 ps---both indicative of high carrier mobility. 

The measured changes in cavity phase shift and SAW amplitude can be converted to changes in the conductivity of graphene at $q=2 \pi/(20~ {\rm \mu m})$ and $\omega=2 \pi\cdot 200~ {\rm MHz}$. In GaAs/AlGaAs heterostructures, this conversion is performed using \autoref{eq:conductivity} with a simple relaxation-type model and the results agree well with experiment \cite{wixforth_surface_1989}. 


To perform this conversion on our data, we perform finite-element simulations of the full resonator and heterostructure assembly, including the graphite top gate that which also contributes to the cavity phase shift (see Supplementary Information). We used published conductivity data from the quantum Hall regime of graphene to estimate the change in cavity phase shift between Landau levels (see Supplementary Information) \cite{kim2019even}. Our model predicts a phase shift of the same order of magnitude that we observe in our data. The extraction of the conductivity from the SAW resonator data could be made quantitative with two improvements: 1) by greatly reducing the top-gate conductance by switching to a low-conductivity material such as MoS$_2$; and 2) by increasing the sample size and/or reducing the total cavity area.

We used SAW resonators operating at 200 MHz in this study because the 20 $\mu$m feature size is easily achievable with cost-efficient, standard photolithographic techniques. SAW resonators operating up into the GHz frequency range, however, are a well-established and mature technology---a typical smartphone contains more than a dozen SAW filters and delay lines. Having demonstrated that LiNbO$_3$ SAW resonators have the requisite sensitivity for small heterostructures, it should now be possible to explore wave-vector dependent effects down to hundreds-of-nanometer length scales. Immediately accessible experiments based on existing theoretical proposals include searching for the crossover from Dirac to Schrodinger-like behaviour in the longitudinal conductivity of graphene \cite{thalmeier2010surface}, and measuring the SAW attenuation of twisted bilayer graphene as a function of chemical potential to look for non-Fermi-liquid signatures \cite{kumar2020fractal}. With further development, one could imagine using SAW resonators to impose dynamic, periodic potentials on vdW heterostrucures that are both tunable in period and switchable on nanosecond timescales.

\section{Acknowledgments}
The fabrication of vdW heterostructures on SAW devices was supported by the Cornell Center for Materials Research with funding from the NSF MRSEC program (DMR-1719875). B.J.R. and Y.F. acknowledge funding from the National Science Foundation under grant no. DMR-1752784.

\end{document}